\documentclass[final,5p,times,twocolumn]{elsarticle}

\usepackage[hidelinks]{hyperref}
\usepackage[T1]{fontenc}
\usepackage[utf8]{inputenc}
\usepackage[british]{babel}
\usepackage{graphicx} % Required for inserting images
\usepackage{amsmath}   
\usepackage{xcolor}
\usepackage{xspace}

\newcommand{\fas}{FASER$\nu$\xspace}
\newcommand{\snd}{SND@LHC\xspace}
\newcommand{\pp}{$pp$\xspace}
\newcommand{\fbinv}{fb$^{-1}$\xspace}
\newcommand{\mum}{$\mu$m\xspace}
\newcommand{\nue}{\ensuremath{\nu_e}\xspace}
\newcommand{\numu}{\ensuremath{\nu_{\mu}}\xspace}
\newcommand{\nutau}{\ensuremath{\nu_{\tau}}\xspace}

\newcommand{\numubar}{\ensuremath{\bar{\nu}_{\mu}}\xspace}
\newcommand{\nutaubar}{\ensuremath{\bar{\nu}_{\tau}}\xspace}

\begin{document}

\begin{frontmatter}

%% Note: \pmbanner before the actual title
\title{A roadmap for neutrino detection at LHC, HL-LHC and SPS}

\author[1,2,3]{Elena Graverini\corref{cor1}}\ead{elena.graverini@cern.ch}
 \cortext[cor1]{Corresponding author}
 \author{on behalf of the SND@LHC and SHiP Collaborations}
\affiliation[1]{organization={Universit\`{a} di Pisa}, 
                 %addressline={},
                 %postcode={}, 
                 city={Pisa}, 
                 country={Italy}}
\affiliation[2]{organization={INFN Sezione di Pisa}, 
                city={Pisa}, 
                country={Italy}}
\affiliation[3]{organization={\'{E}cole Polytechnique F\'{e}d\'{e}rale}, 
                city={Lausanne}, 
                country={Switzerland}}
\begin{abstract}
SND@LHC is a new detector for neutrino physics at LHC. Its experimental configuration makes it possible to distinguish between all three neutrino flavours, opening a unique opportunity to probe physics of heavy flavour production at the LHC in the region that is not accessible to ATLAS, CMS and LHCb. It can also explore lepton flavour universality in the neutral sector, and search for feebly interacting particles. The detector has been commissioned and installed in 2021-2022. A first set of data has since then been collected, providing the first observation of neutrinos produced at a collider. This paper discusses the detector technologies being used to study high-energy neutrinos at the LHC, and their performance in terms of physics reach. The necessary upgrades to operate at high-luminosity LHC are presented, as well as a proposed experiment to perform neutrino measurements at the newly approved Beam Dump Facility.
\end{abstract}

\begin{keyword}
SND@LHC Neutrino LHC SHiP BDF
\end{keyword}

\end{frontmatter}

%text of the article
\section{Introduction}\label{sec:intro}
Interest in high-energy neutrinos produced at \pp colliders dates back to the '80s~\cite{DeRujula:1984pg}. Neutrinos are produced at the CERN Large Hadron Collider (LHC) in leptonic $W$ and $Z$ decays, as well as in decays of hadrons containing heavy $b$ or $c$ quarks. Subsequent decays of lighter secondary particles (pions and kaons) also produce neutrinos.

What makes the LHC unique as a neutrino factory is the large energy of collisions. Measurements with neutrinos have historically been performed at low energy, for oscillation studies, reaching an energy of 350~GeV only for muon neutrinos. Very high-energy cosmic rays provide a source of neutrinos above 10~TeV, while the range 350~GeV--10~TeV remains unexplored~\cite{ParticleDataGroup:2022pth, Bustamante:2017xuy}.

Only in the last decade, however, was it shown that backgrounds at the LHC are low enough for neutrino experiments~\cite{Buontempo:2018gta,Beni:2019gxv,Beni:2020yfy}. At LHC energies, neutrino interaction cross-sections are relatively large, and the high flux of neutrinos expected in the forward direction implies that a relatively small compact detector would have significant potential for physics.

\section{The \snd experiment}\label{sec:snd}
\snd~\cite{SNDLHC:2022ihg} is a new experiment conceived to study neutrinos of all flavours at LHC. It has been installed and commissioned in 2021-2022, and it has been taking data since the start of the LHC Run~3. It is expected to collect an integrated luminosity of 250~\fbinv by 2025, corresponding to roughly 2000 neutrino interactions.

The \snd detector is located in the TI18 service tunnel, on the path of the flux of high-energy neutrinos produced in \pp collision at the ATLAS interaction point IP1. It looks in the opposite direction of IP1 with respect to the only other neutrino detector currently in operation at LHC, \fas~\cite{FASER:2019dxq}. \snd is located slightly off-axis from the neutrino beam, with pseudorapidity coverage $7.2<\eta<8.4$, which allows it to probe neutrinos which largely originate from decays of charmed mesons. This pseudorapidity region cannot be covered by the any of the four large LHC experiments.

\begin{figure}
\hspace{-5mm}\includegraphics[width=1.1\linewidth]{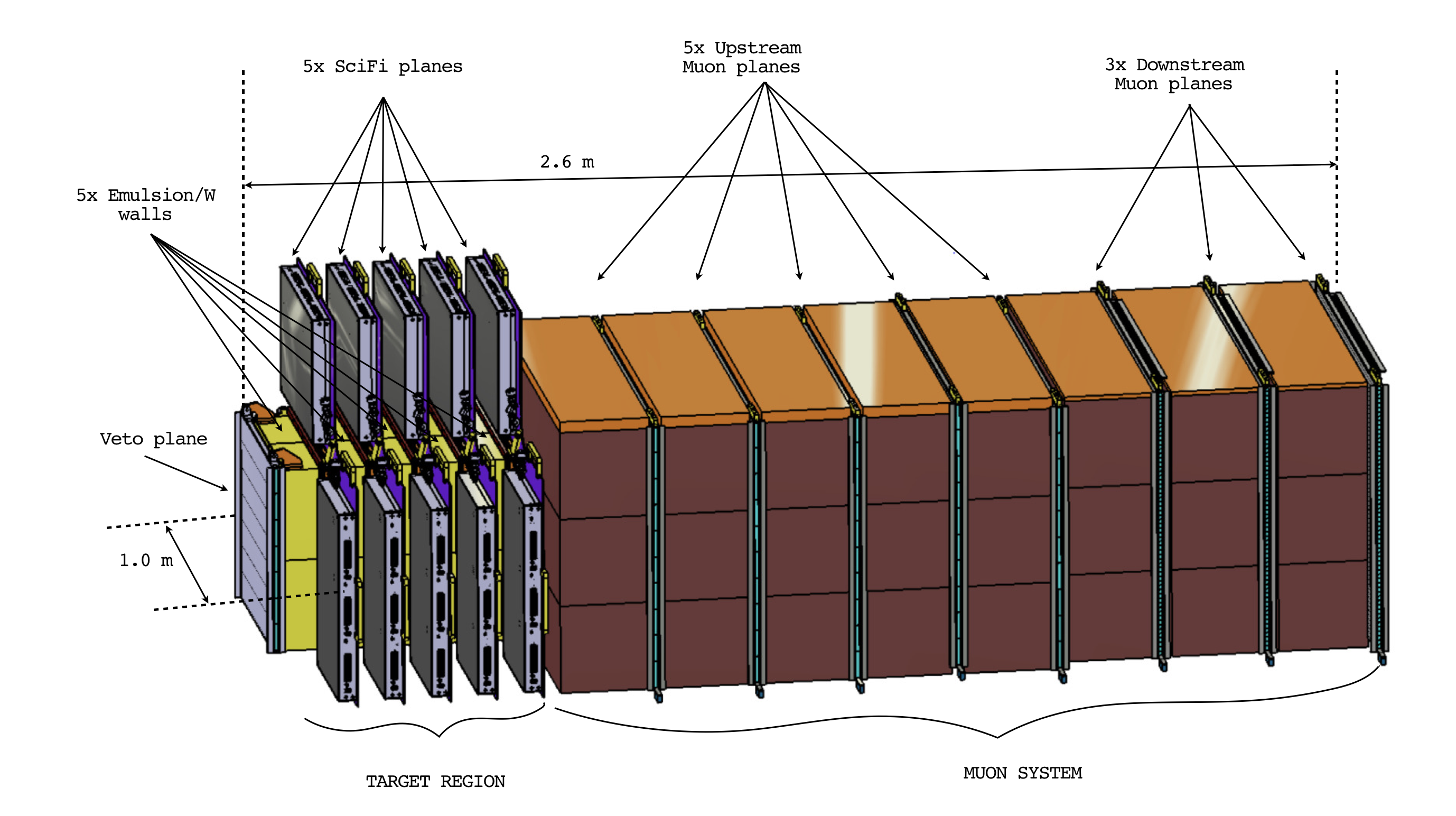}
\caption{Layout of the SND@LHC experiment~\cite{SND:TP}.}\label{fig:snd_layout}
\end{figure}

The TI18 tunnel is located about 480~m away from IP1, which shields \snd from most charged particles produced in \pp collisions. The detector is composed of four parts, schematised in \figurename~\ref{fig:snd_layout}. Events where a charged particle has entered the detector from the front (mostly consisting of muons) are removed using a recently upgraded veto detector~\cite{giulia}, composed of three orthogonal planes of $1\times 6\times 42$~cm$^3$ scintillating bars, read out by Hamamatsu S14160-6050HS silicon photomultipliers (SiPM) at both ends. Each bar is made of EJ-200 scintillator and isolated by a cover made of Mylar foils. After the upgrade (addition of a third plane) the veto system reached an inefficiency of $10^{-8}$, measured in 2024 data.

The target and vertex detector consists of 830~kg of Emulsion Cloud Chambers (ECC)~\cite{Acquafredda:2009zz} and it is further instrumented with an electronic tracker. Each of the five walls comprises 4 ECC bricks, each made of 60 emulsion films alternated with 1~mm thick tungsten plates acting as passive material. Emulsion films are replaced approximately every 25~\fbinv during short technical stops, with a procedure requiring an access of at most 5~hours, and they are analysed by a newly-developed fast procedure using fully automated optical microscopes~\cite{Alexandrov:2015kzs}. The replacement rate has been calibrated based on a measurement of the muon flux in the TI18 tunnel~\cite{SNDLHC:2023mib}.
Five scintillating fibre (SciFi)~\cite{LHCb:2014uqj} planes are interleaved to the ECC walls. These electronics trackers have three functions: (i) providing a time-stamp to neutrino interactions reconstructed in the ECCs; (ii) performing a real-time energy measurement of electromagnetic showers, in sampling calorimeter setup that uses the ECCs as passive material; and (iii) measuring the energy of hadronic showers in combination with the scintillating bars of the muon detector. Each scintillating fibre mat is composed of 6 staggered layers of double-cladded polystyrene fibres with a 250~\mum diameter, glued together with titan oxide-loaded epoxy glue in order to prevent cross-talk between fibres, and read at one end by an array of Hamamatsu S13552 SiPMs. While this technology is already in use at the LHCb experiment, the readout electronics have been reoptimised to increase time resolution and detect electromagnetic showers.

Eight layers of scintillating bars interleaved with 20~cm thick iron slabs, placed downstream of the neutrino target, act as sampling hadronic calorimeter, and identify muons passing through the detector. The total thickness is about 11~$\lambda_{int}$ for showers starting in the target region. The first five layers use 6~cm thick EJ-200 scintillating bars, while the last three layers have a higher granularity geometry conceived in view of muon identification, and use alternating layers of horizontal and vertical 1~cm thick bars. The whole system is read out at both ends by Hamamatsu S14160 SiPMs, with the exception of the vertical planes of the downstream layers, as the bars are located as close as possible to the cavern floor to improve acceptance.  

The whole detector is surrounded by a 30\% borated polyethylene and acrylic box acting as once as a low-energy neutron shield and as an insulation chamber with controlled temperature and humidity, which is necessary for the correct operation of the emulsion films.

All subsystems are read out with the same electronics, consisting of TOFPET2 ASIC front-end boards and custom DAQ boards based on the Mercury SA1 module from Enclustra~\cite{SNDLHC:2022ihg}.

DPMJET3, FLUKA, \textsc{Genie} and \textsc{Geant}4 are used to perform simulation studies. About 550 neutral-current and three times as many charged-current interactions are expected in the detector, from neutrinos of all three flavours, in a mixture of about 72\% \numu, 23\% \nue and 5\% \nutau without distinction between neutrinos and anti-neutrinos. The main physics case of the current stage of \snd is the measurement of charm production using \nue, a measurement that can be performed with a statistical uncertainty smaller than 5\% and that is dominated by systematic uncertainties of about 35\%. Reducing these systematic uncertainties led the collaboration to propose an upgrade of the experiment, discussed in Section~\ref{sec:adv}. This measurement can provide insight into the $gg\to c\bar{c}$ scattering process, with \snd capable of probing a gluon momentum fraction $x$ ranging down to $10^{-6}$, where the gluon PDF is completely unknown. Measuring the latter at such low $x$ can provide crucial input for future experiments with a similar sensitivity range, such as FCC~\cite{FCC:2018byv}. The data collected at \snd will furthermore enable lepton flavour universality (LFU) tests in the neutrino sector. With the LHC Run~3 data sample, these tests can reach about 30\% statistical and 20\% systematic uncertainty in the $\nue/\nutau$ ratio, dominated by the low yield of \nutau, and 10\% statistical and 10\% systematic uncertainty in the $\nue/\numu$ ratio when restricted to $E_{\nu}>600$~GeV to exclude contamination from pion and kaon decays.

\begin{figure}
    \centering
    \includegraphics[width=.8\linewidth]{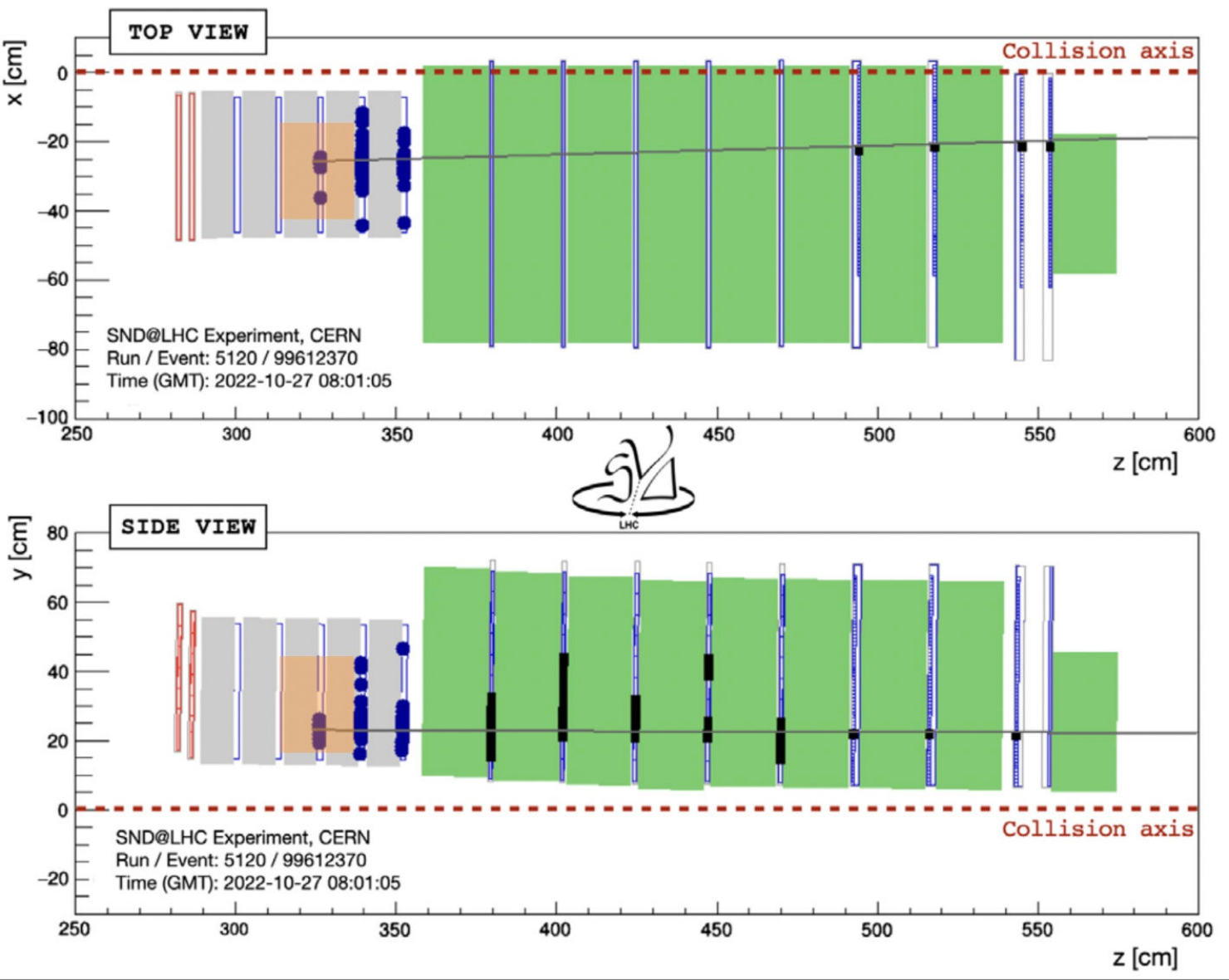}
    \caption{Top and side view of a recorded \numu CC scattering event at \snd~\cite{SNDLHC:2023pun}.}
    \label{fig:ev-disp}
\end{figure}

\snd can also perform model-independent searches for feebly interacting particles (FIPs) in scattering signatures~\cite{Boyarsky:2021moj}. The main background for direct searches is due to neutrino interactions, and it can be vastly reduced by measuring the time-of-flight (TOF), which is enabled by the 200~ps resolution of the SciFi detector. The significance of the search will therefore depend on the FIP mass. New particles may also be searched for in decay signatures, thanks to the excellent spatial resolution provided by the ECC technology.

In early 2023, both \fas and \snd collaborations published their first observation of collider muon neutrinos, at about 16$\sigma$ and 6.8$\sigma$, respectively, using data collected in 2022~\cite{FASER:2023zcr,SNDLHC:2023pun}. One of the candidates observed by \snd is shown in \figurename~\ref{fig:ev-disp}. The \snd collaboration recently updated the analysis with 2023 data and expanded the fiducial volume, raising the amount of collected neutrino candidate interactions from 8 to 32 and the significance to 12$\sigma$, while also providing a first measurement of kinematic spectra and a first observation of \nue-like events based solely on the electronic detectors~\cite{cv-moriond2024}. Analysis of the emulsion films is ongoing.

\section{High-luminosity upgrade: AdvSND}\label{sec:adv}
\begin{figure}
    \centering\includegraphics[width=\linewidth]{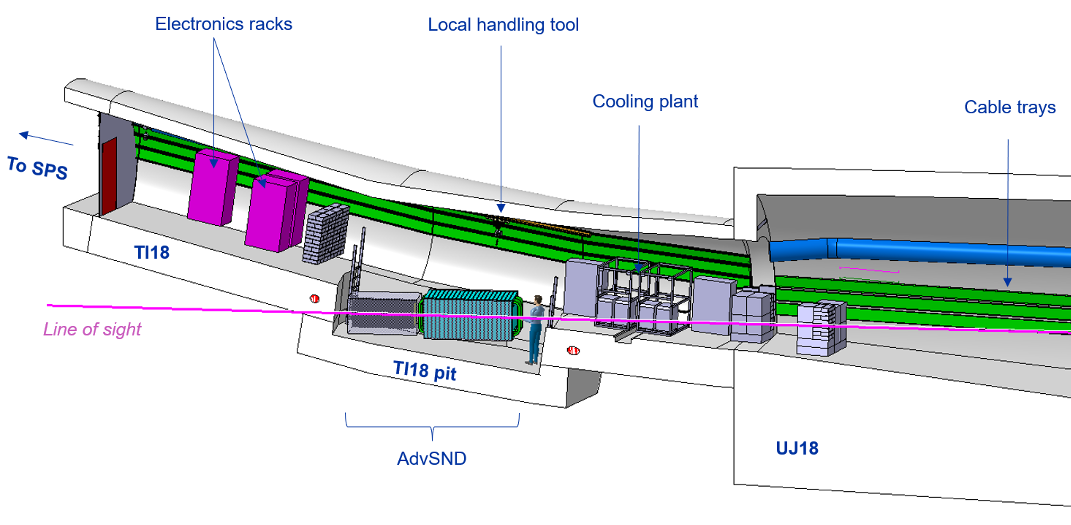}
    \caption{Layout of the AdvSND detector including civil engineering in the TI18 LHC service tunnel~\cite{lhcc_focus}.}\label{fig:advsnd}
\end{figure}
An upgrade of the detector, AdvSND, is proposed for LHC Run~4, in order to fully exploit the potential of the future High-Luminosity LHC (HL-LHC) collider upgrade~\cite{AdvSND:LoI}.
The use of nuclear emulsions will not be an option, given the foreseen five-fold increase in instantaneous luminosity. Furthermore, the collaboration seeks to implement structural design changes that had been impossible for Run~3 given the short time available for detector installation. In fact, the current configuration of TI18 (sloping floor and available space) has imposed severe limitations on the \snd acceptance and performance.

Civil engineering interventions are proposed in order to enhance the detector acceptance and to accommodate a magnetised hadronic calorimeter. The latter would allow for a first observation of tau anti-neutrinos, the last particle of the SM that is known to exist but has not yet been observed. The foreseen layout for AdvSND is undergoing intense studies and it is schematised in \figurename~\ref{fig:advsnd}.

\begin{figure}
    \centering
    \includegraphics[width=.8\linewidth]{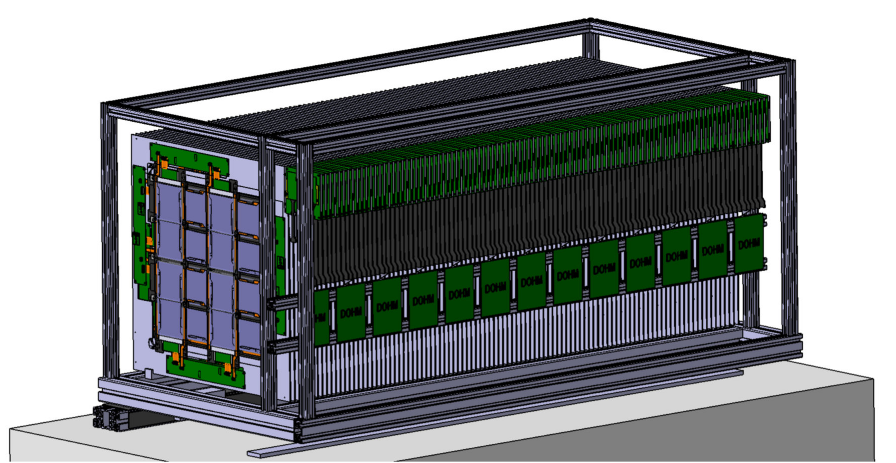}
    \caption{View of the full stack of 100 tungsten-silicon tracker stations for AdvSND~\cite{AdvSND:LoI}.}
    \label{fig:AdvSND-target}
\end{figure}

It is proposed to reuse the silicon microstrip sensors from the two outermost layers of the CMS Tracker Outer Barrel (TOB)~\cite{cms_tracker} to build the new AdvSND target and electromagnetic calorimeter. The CMS board has approved this transition on February 9, 2024. Eight $10\times 20$~cm$^2$ modules will be mapped to one $40\times 40$~cm$^2$ SND target station. The target itself will consist of 100 alternating layers of tungsten absorbers and silicon trackers, as seen in \figurename~\ref{fig:AdvSND-target}. A prototype is under construction and its performance will be tested in summer 2024. An option is being considered, to replace 50 layers with large-scale Monolithic Stitched Sensors (MOSS) in synergy with the ALICE ITS3 upgrade. These sensors could overlap each other in the central region and thus provide tracklets in a way similar to what was achieved with emulsion films.

A dual system comprising a near and a far detector is a possibility being explored for the LHC Run~5. AdvSND, as described above, will act as a far detector. A near detector could be placed in the UJ57/56 hall, near CMS. This detector would overlap with the LHCb acceptance, where charm and beauty measurements are best performed. The role of the near detector would be to reduce systematic uncertainties on the measurement of the gluon PDF, and to enable a standalone measurement of neutrino cross-sections. Including the near detector, the systematic uncertainty on the gluon PDF measurement can be reduced to the same level of the expected statistical sensitivity, \textit{i.e.} about 1\%. The LFU measurements will also profit from the larger data sample, decreasing the \nue / \nutau statistical uncertainty to 5\% and that of \nue / \numu to a few \%.

\section{The SHiP experiment at the Beam Dump Facility}\label{sec:ship}
\begin{figure}
\centering\includegraphics[width=\linewidth]{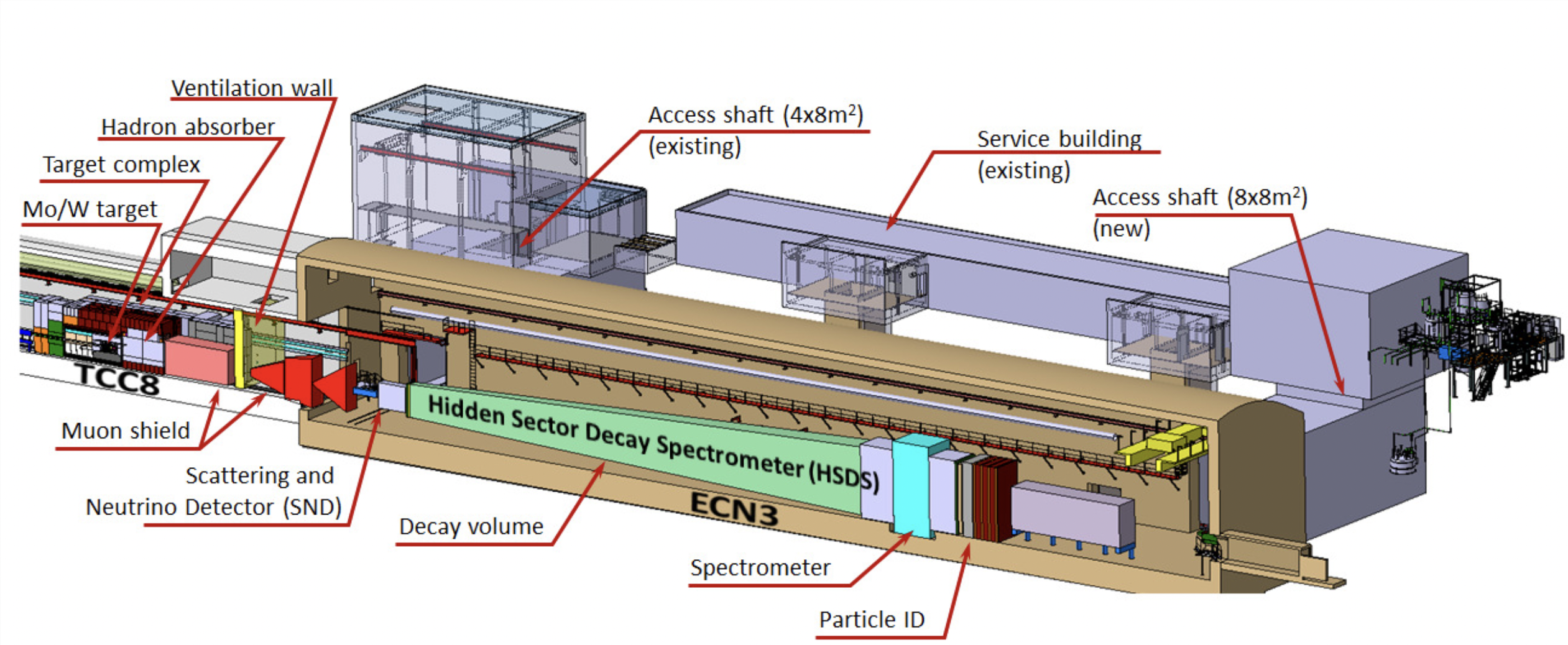}
\caption{Scheme depicting the layout of the SHiP proposed experiment in the ECN3 experimental hall~\cite{SHiP:ECN3}.}\label{fig:ship_ecn3}
\end{figure}
The Beam Dump Facility (BDF) and its main user, the SHiP experiment (Search for Hidden Particles), have recently been approved by the CERN directorate for construction in the North Area. The experiment had first been proposed in 2014 and its design has since shifted in order to make future use of the ECN3 experimental hall currently used by the NA62 experiment~\cite{SHiP:ECN3}. In fact, the \snd detector idea was initially devised for use at the BDF, and it has then been extended for higher-energy operation at LHC.

The SHiP experiment (\figurename~\ref{fig:ship_ecn3}) consists of two parts: a hidden sector decay spectrometer (HSDS), aimed at the direct detection of FIPs, and of a scattering and neutrino detector (SND) for neutrino and light dark matter physics. The BDF will deliver a 400~GeV/$c$ proton beam extracted from the SPS onto a target composed of blocks of titanium--zirconium-doped molybdenum alloy (TZM), cladded by a tantalum-alloy, followed by blocks of tantalum-cladded
pure tungsten. This design optimises the production of heavy-flavoured hadrons and photons, while suppressing light-flavoured hadron decays into muons and neutrinos. This provides the cleanest possible environment for FIP searches. The HSDS detector is composed of a long vacuum vessel serving as fiducial decay volume, preceded and surrounded by veto taggers, and followed by a straw spectrometer, a timing detector and a calorimeter.

A magnetic muon shield precedes the HSDS. Muons are the only particles that emerge from the BDF and constitute background for FIP searches, and their large yield makes the operation of the SHiP detector problematic. An alternate-polarity magnet is designed, to split the muon beam into well separated positively and negatively charged components with a virtually background-free area in the centre, where the SHiP detector is located. This design was re-optimised for ECN3 using deep learning models~\cite{Baranov:2017chy}, and it comprises a first superconducting magnet followed by three normal-conducting magnets.

The SND detector is placed immediately downstream of the muon shield. It comprises a neutrino interaction target using the same technology as the current \snd experiment, and a muon spectrometer designed to identify \numu / \numubar and \nutau / \nutaubar. An option to integrate the neutrino target inside the last elements of the muon shield, to improve the acceptance of the HSDS by pulling it closer to the BDF, is currently being investigated.

The high-intensity beam of the SPS will provide the BDF/SHiP experiment with an unprecedented neutrino yield. About $14\times 10^6$ charged-current interactions are expected in the detector. In particular, the collaboration expects to reconstruct about 53000 tau-flavoured neutrinos. Charge separation can be performed for leptonically-decaying $\tau$ leptons, allowing to correctly identify about 4000 \nutau and 3000 \nutaubar interactions. It should be noted that such a dataset can be used to constrain the \nutau magnetic moment down to about $9\times 10^{-8}\mu_B$. More than $6\times 10^5$ neutrino-induced charmed hadrons are expected, exceeding the samples available in previous experiments by over an order of magnitude~\cite{SHiP:ECN3}.

\section{Conclusions}\label{sec:concl}
A roadmap for neutrino studies at current and future experimental facilities is presented. The ``Scattering and Neutrino Detector'' (SND) concept has been successfully implemented in the \snd detector and it has already provided the first results using the beam of forward neutrinos produced in \pp interactions at the LHC IP1. Such detector is also capable of dark matter searches in scattering and decay signatures.
A proposed upgrade in view of high-luminosity LHC is presented, aimed at greatly reducing the main systematics for the measurement of the gluon PDF in $pp\to c\bar{c}$ scattering.
Finally, a renewed implementation of the original SND detector at the future Beam Dump Facility is reviewed. This detector will profit from unprecedented statistics for all neutrino flavours.

\section*{Acknowledgements}
The author gratefully acknowledges support from the Swiss National Science foundation, grant number PZ00P2\_202065.

  \bibliographystyle{elsarticle-num-names} 
  \bibliography{bib}
\end{document}